\documentclass[fleqn,usenatbib,letters]{mnras}

\usepackage[T1]{fontenc}
\usepackage{ae,aecompl}
\usepackage{graphicx}	% Including figure files
\usepackage{amsmath}	% Advanced maths commands
\usepackage{amssymb}	% Extra maths symbols
\usepackage{interval}  %for prior intervals
\usepackage{relsize}
\usepackage{balance}

\usepackage[usenames,dvipsnames]{xcolor}

\hypersetup{
    colorlinks = true,
    citecolor = {MidnightBlue},
    linkcolor = {BrickRed},
    urlcolor = {BrickRed}
}

 \newcommand{\plus}{ + }
 \newcommand{\minus}{ - }

 \newcommand{\tvspace}{\vspace*{4.pt}}

\newcommand{\omegak}{\Omega_{\rm K}}
\newcommand{\Nde}{N^\mathrm{DE}_{\rm bins}}

 \title[Model-independent curvature determination]{Model-independent curvature determination with \\ 21cm intensity mapping experiments}
\author[Witzemann et al.]{
Amadeus Witzemann,$^{1,2}$\thanks{E-mail: amadeus.witzemann@gmx.at}
Philip Bull,$^{3,4}$
Chris Clarkson,$^{5,1,2}$
\newauthor{
Mario G. Santos,$^{1,6}$
Marta Spinelli$^{1}$
and Amanda Weltman,$^{2,7}$
}
\\ \\
$^{1}$Department of Physics and Astronomy, University of the Western Cape, Cape Town 7535, South Africa \\
$^{2}$Department of Mathematics and Applied
Mathematics, University of Cape Town, Rondebosch 7701, South Africa \\
$^{3}$Department of Astronomy, University of California Berkeley, Berkeley, CA 94720, USA \\
$^{4}$Radio Astronomy Laboratory, University of California Berkeley, Berkeley, CA 94720, USA \\
$^{5}$School of Physics and Astronomy, Queen Mary, University of London, Mile End, London, UK\\
$^{6}$SKA South Africa, 3rd Floor, The Park, Park Road, Pinelands, 7405, South Africa\\
$^{7}$School of Natural Sciences, Institute for Advanced Study, Olden Lane, Princeton, NJ, 08540, USA
}

\date{Accepted XXX. Received YYY; in original form ZZZ}

\pubyear{2017}

\begin{document}
\label{firstpage}
\pagerange{\pageref{firstpage}--\pageref{lastpage}}
\maketitle

% Abstract of the paper
\begin{abstract}
Recent precision cosmological parameter constraints imply that the spatial curvature of the Universe is essentially dynamically negligible~-- but only if relatively strong assumptions are made about the equation of state of dark energy (DE). When these assumptions are relaxed, strong degeneracies arise that make it hard to disentangle DE and curvature, degrading the constraints.
We show that forthcoming 21cm intensity mapping experiments such as HIRAX are ideally designed to carry out model-independent curvature measurements, as they can measure the clustering signal at high redshift with sufficient precision to break many of the degeneracies. We consider two different model-independent methods, based on `avoiding' the DE-dominated regime and non-parametric modelling of the DE equation of state respectively. Our forecasts show that HIRAX will be able to improve upon current model-independent constraints by around an order of magnitude, reaching percent-level accuracy even when an arbitrary DE equation of state is assumed. In the same model-independent analysis, the sample variance limit for a similar survey is another order of magnitude better.
\end{abstract}

\begin{keywords}
dark energy --  large-scale structure of Universe -- cosmology: observations
\end{keywords}

%%%%%%%%%%%%%%%%%%%%%%%%%%%%%%%%%%%%%%%%%%%%%%%%%%

\section{Introduction}

Most viable models predict that only a very small amount of curvature should remain after the end of inflation -- smaller even than the Hubble-scale curvature perturbations generated by quantum fluctuations. While some theories can generate observable amounts of curvature, they tend to either be somewhat contrived \citep[e.g. see][]{1995PhRvD..52.3314B, 1996PhRvL..77..215C}, or are likely to be accompanied by large-scale anomalies that would be visible in the CMB \citep{2015PhRvD..91l3523A}. Furthermore, several major classes of inflationary theories explicitly predict that curvature should be small. False vacuum eternal inflation would be ruled out by a detection of the curvature density parameter at the $\omegak < -10^{-4}$ level, for example, while slow-roll eternal inflation would be ruled out if $\omegak > +10^{-4}$ \citep{Kleban2012, 2012PhRvD..86b3534G}.

Recent cosmological parameter constraints, most notably from the Planck CMB temperature and polarisation spectra combined with baryon acoustic oscillation (BAO) constraints \citep{2016A&A...594A..13P}, have placed upper limits on curvature of $|\omegak| < 5 \times 10^{-3}$ (95\% CL) -- still a factor of 50 in precision away from being able to put any serious pressure on eternal inflation \citep{Vardanyan:2009ft, Leonard:2016evk}. However, this figure is only achieved after making strong assumptions about the nature of dark energy (DE), i.e. that it behaves exactly like a cosmological constant, with an equation of state of $w = -1$. In fact, the equation of state of DE remains unknown, with many candidate theories predicting slightly different equations of state that can vary substantially with redshift \citep[e.g.][]{Huterer:2006mv, Marsh:2014xoa, Raveri:2017qvt}.

When $w$ is allowed to vary, the uncertainty increases on all parameters, as the data must now constrain several additional degrees of freedom. A common choice of parametrization, $w(a) \approx w_0 + w_a (1 - a)$, introduces only two additional degrees of freedom, but more general `non-parametric' analyses \citep[e.g. see][]{2014PhRvD..90f3006N} can introduce many more. Many observables also depend on combinations of cosmological functions, like the Hubble rate, for which there is at least a partial degeneracy between $w(z)$ and $\omegak$ (since, at the background level, an arbitrary DE equation of state can partially mimic the redshift scaling of curvature). This goes beyond the well-known `geometric degeneracy', in which $\omegak$ and $\Omega_{\rm DE}$ are degenerate when constrained by the primary CMB power spectrum alone -- even probes that constrain distances or the Hubble rate at multiple redshifts are susceptible to some (typically strong) degree of correlation between $\omegak$ and $w$~\citep{2007JCAP...08..011C,2008GReGr..40..285H}. The degeneracies that exist in distances and the Hubble rate pull in opposing directions however, implying that a combined measurement, using the BAO feature or similar, can reduce the degeneracy significantly -- even with no assumptions on $w(z)$, as we show here \citep[see also][]{Takada:2015mma}.

It is also possible to sidestep the problem of modelling $w(z)$ entirely, if one is willing to make a relatively mild assumption about the nature of DE. If the energy density of DE becomes negligible at some point in the past, i.e. $\Omega_{\rm DE}(z) \to 0$ beyond some $z > z_M$, it is possible to construct combinations of distance measures such that the DE-dependent part cancels out \citep{2006PhRvD..73b3503K}. In principle, this results in an observable that depends only on the matter and curvature contributions to the Friedmann equation at $z > z_M$, sufficiently deep into the matter-dominated regime. For typical values of cosmological parameters, matter domination occurs at $z \gtrsim 2$, and so only high-redshift probes such as the Lyman-alpha forest or 21cm intensity mapping (IM) can be used for this test.

21cm intensity mapping is a relatively new technique that measures the combined 21cm spectral line emission from many unresolved galaxies in each pixel \citep{2017arXiv170909066K}. By trading angular resolution for spectral resolution and sensitivity, one can rapidly survey large cosmological volumes while retaining most cosmological information on large scales. Since the underlying galaxy distribution is a biased tracer of the cosmic matter distribution, so too are the measured intensity maps. By using these in a similar way to other galaxy clustering observables \citep[e.g. for BAO measurements:][]{Chang:2007xk, Bull:2015nra, Villaescusa-Navarro:2016kbz}, one can measure distances out to significantly higher redshift than a typical optical galaxy survey \citep{2015ApJ...803...21B}. This is especially true of 21cm IM, which uses the 21cm line from neutral hydrogen (HI) as a tracer. Since HI is ubiquitous in the Universe out to relatively high redshift, and since radio telescopes can readily be built to cover very large frequency (and thus redshift) ranges, 21cm IM is well-suited to performing large, high-redshift cosmological surveys -- and thus testing curvature in a model-independent way.

In this Letter, we study the ability of forthcoming 21cm IM experiments to constrain curvature in a model-independent way. Our particular focus is on the {\sl Hydrogen Intensity and Real-time Analysis eXperiment} \citep[HIRAX;][]{2016SPIE.9906E..5XN}, a radio interferometer currently under construction in South Africa. We assume a Planck 2015 flat $\Lambda$CDM fiducial cosmology throughout, with $\omegak=0$, $\Omega_\mathrm M = 0.316$, $\Omega_b = 0.049$, $\Omega_{\rm rad} = 9.13 \times 10 ^{-5}$, and $h=0.67$ \citep{2016A&A...594A..13P}.

%%%%%%%%%%%%%%%%%%%%%%%%%%%%%%%%%%%%%%%%%%%%%%%%%%

\section{Curvature measurements in the presence of dark energy}

We now present two different methods for obtaining curvature constraints that are independent of the assumed dark energy model, at least in principle. The first (Sect.~\ref{section:sidestepping}) is based on constructing combinations of observables that do not depend on the low redshift, dark energy-dominated regime. The second (Sect.~\ref{section:pwcw}) uses a non-parametric approach to modelling the DE equation of state (EOS), marginalising over its value in many redshift bins to produce curvature constraints that are independent of any assumed functional form for $w(z)$. For comparison, we also compare with a simple, commonly used 2-parameter dark energy model in Sect.~\ref{section:series}.

%%%%%%%%%%%%%%%%%%%%%%%%%%%%%%%%%%%%%%%%%%%%%%%%%%

\subsection{Avoiding the dark energy era}
\label{section:sidestepping}

One way of obtaining model-independent curvature constraints is to try to avoid the DE era entirely. We extend the approach described in \cite{2006PhRvD..73b3503K} to derive a combination of distance measures that is linearised in the spatial curvature, $k \equiv -\omegak H_0^2$, and which is relatively independent of the dark energy contribution to those distances.

In a FLRW universe with line element $ds^2 = -dt^2 + a^2 dr^2 / \sqrt{1 - k r^2}$, the comoving distance is given by $\chi = \int_0^r {dr^\prime}/{\sqrt{1-k{r^\prime}^2}}$.
The series expansion of $r(\chi, k)$ for small $k$ is then
$r \approx \chi - \frac{\chi^3 k}{6}  + \mathcal{O}(k^2),\,$
regardless of the sign of $k$. We identify the coordinate distance $r = \tilde{D}_{A}(\chi, k) = D_A / a$, where $\tilde{D}_A$ is the comoving angular diameter distance, so that $\tilde{D}_A(\chi, k) \approx \chi - {\chi^3 k/6}$ to first order in $k$.
The comoving distance is additive, i.e. $\chi_\mathrm{OL}=\chi_\mathrm{OM}+\chi_\mathrm{ML}$,  where $\chi_\mathrm{OL}$ is the comoving distance from the observer to the last scattering surface (LSS), $\chi_\mathrm{OM}$ is the comoving distance from the observer to an intermediate redshift $z_\mathrm M$, where we assume that dark energy can be neglected, and $\chi_\mathrm{ML}$ is the distance from $z_\mathrm M$ to the LSS.
Solving $\tilde{D}_\mathrm{OL} = \tilde{D}_A(\chi_\mathrm{OL}, k) \approx \chi_\mathrm{OL} - \chi_\mathrm{OL}^3k/6$ for $k$ gives, to first order,
\begin{equation}
\label{eq:knoxk}
k=6\,\biggl(\frac{\tilde{D}_\mathrm{OM}+\chi_\mathrm{ML}-\tilde{D}_\mathrm{OL}}{(\tilde{D}_\mathrm{OL})^3-(\tilde{D}_\mathrm{OM})^3}\biggr),
\end{equation}
where $\tilde{D}_\mathrm{OM}$ and $\tilde{D}_\mathrm{OL}$ denote the comoving angular diameter distances from the observer to $z_\mathrm M$ and the LSS respectively. Both are directly observable, but $\chi_{\rm ML}$ is not. (Note that Eq.~\ref{eq:knoxk} differs from the \cite{2006PhRvD..73b3503K} result by a minus sign.)

\begin{figure}
\vspace{-1em}
\hspace{-1em}
\includegraphics[width=1.08\columnwidth]{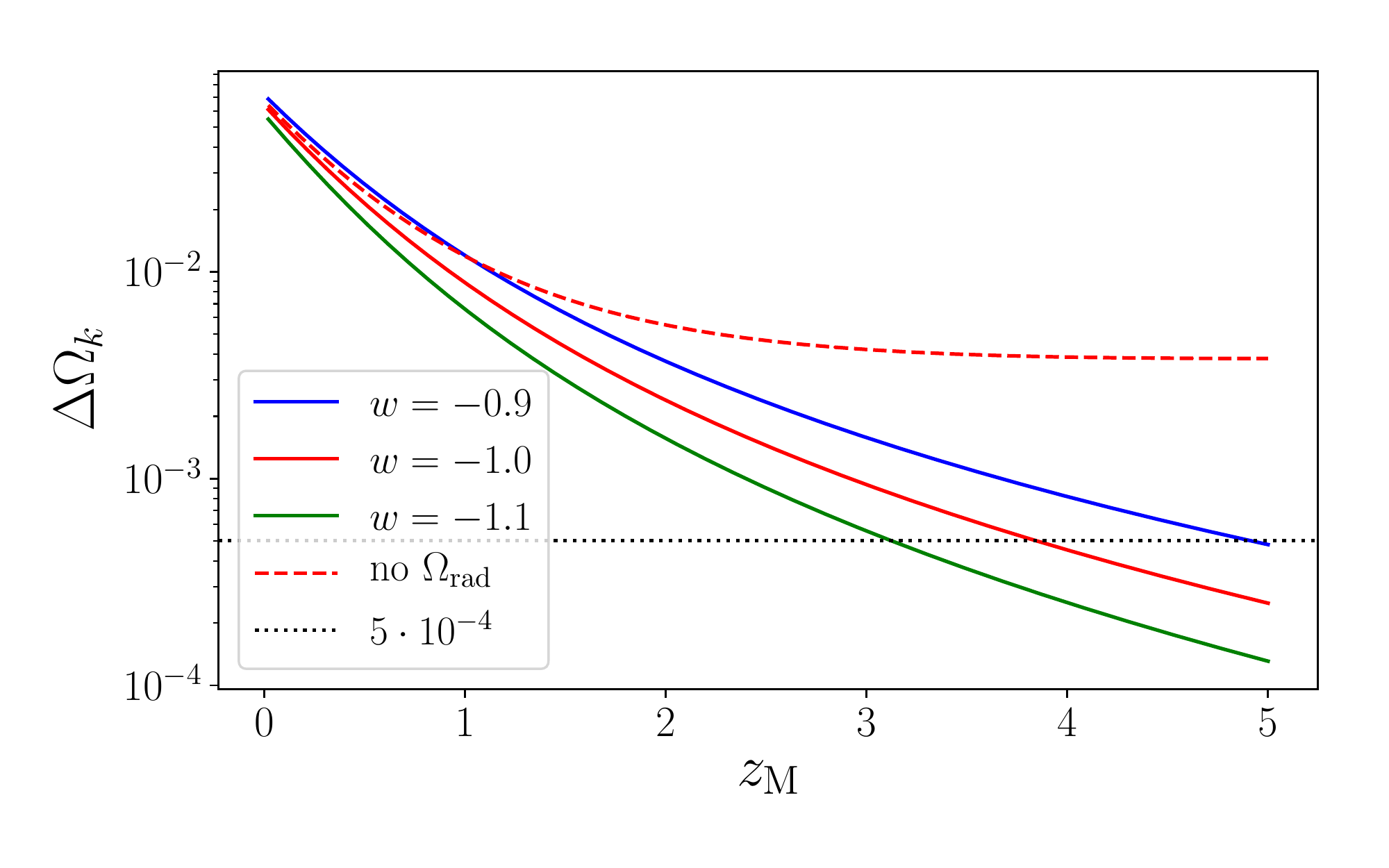}
\vspace{-1.em}
\caption{Bias on the recovered value of $\omegak$ using the avoidance method, as a function of minimum redshift, $z_M$, for several input values of $w$. The red dashed line shows the behaviour of the original method from \citet{2006PhRvD..73b3503K}, which neglected radiation, for $w=-1.0$. The dotted line shows roughly the level of the `curvature floor', or minimum observable curvature \citep{Vardanyan:2009ft, Leonard:2016evk}.}
\label{fig:kbias}
\end{figure}

To obtain curvature constraints using this method, we use the \cite{2016A&A...594A..13P} measurement of $\tilde{D}_{OL}$, and the HIRAX forecasts or SDSS measurements \citep{2017MNRAS.464.1493S} of $\tilde{D}_\mathrm{OM}$, plus simple error propagation, to estimate the error on $k$. We can then approximate $\chi_{\rm ML}$ by neglecting curvature and DE for $z > z_M$, to give
\begin{eqnarray}
\label{lmleq}
\chi_\mathrm{ML} &\approx& H_0\int_{z_\mathrm M}^{z_*} dz/\sqrt{\Omega_\mathrm M (1+z)^3 + \Omega_\mathrm{rad} (1+z)^4} \nonumber \\
 &=& \frac{-2}{H_0 \Omega_\mathrm M} \sqrt{\frac{\Omega_\mathrm M}{1+z} + \Omega_\mathrm{rad}}\Big|_{z_\mathrm{M}}^{z_*} \equiv \chi^\mathrm{md}_\mathrm{ML}(z_\mathrm M),
\end{eqnarray}
where $\Omega_{\rm rad}$ is the fractional energy density in radiation, including photons and neutrinos. The corresponding expression in \cite{2006PhRvD..73b3503K} also neglected radiation, but this would bias $\omegak$ at around the $10^{-2}$ level, as shown in Fig.~\ref{fig:kbias}. For models close to $\Lambda$CDM, the relative difference between $\chi^\mathrm{md}_\mathrm{ML}$ and the true $\chi_\mathrm{ML}$ (including DE and curvature) quickly drops below $10^{-2}$ for $z_M \gtrsim 1.5$. The bias in $\omegak$ for a handful of different values of $w$ are also shown in Fig.~\ref{fig:kbias}. The implication is that curvature measurements made only at higher redshifts are much less sensitive to the detailed DE behaviour, although the choice of minimum redshift will depend on the target precision on $\omegak$. For example, to ensure a bias below $\Delta \omegak \approx 10^{-2}$ and $10^{-3}$ for a reasonable spread of $w$ values, one would take $z_M \gtrsim 1.3$ and $\gtrsim 4$ respectively.

%%%%%%%%%%%%%%%%%%%%%%%%%%%%%%%%%%%%%%%%%%%%%%%%%%

\subsection{Non-parametric dark energy marginalisation}
\label{section:pwcw}

As shown above, trying to avoid the DE-dominated regime still results in a mild model dependence, as the choice of $z_M$ and the expected bias in $\omegak$ both depend on the DE EOS. There is also the issue that a large amount of low-redshift data must be discarded. In this section we consider an alternative model-independent approach, based on a piecewise constant parametrisation of the EOS. For a sufficiently large number of bins, this allows us to closely approximate essentially any arbitrary EOS. Marginalising over the $w$ values in all bins then produces curvature constraints that are free of any assumptions about the particular form of $w(z)$, at least in principle.

We define a general piecewise EOS by setting $w(z) \equiv w_i$ for $z_i<z<z_{i+1}$, $i = 1\ldots \Nde$, and
choose a binning that is equally spaced in scale factor in this instance.
The fractional dark energy density is then given by
\begin{equation}
  \label{eq:omLFINAL}
  \Omega_\mathrm{DE}(z)=
	\Omega_{\mathrm{DE},0}\, (1+z)^{3(1+w_\eta)} \prod_{i=2}^{\eta} ( 1+z_i)^{3(w_{i-1}-w_i)}\,,
\end{equation}
where $\eta$ is chosen such that $z_\eta \le z < z_{\eta + 1}$ for a given $z$.
%where $ \eta \in \mathbb{N}$ such that $z_\eta = \mathrm{max}\{z_i | z_i < z \}$.
%(When $\eta=1$ the product term is absent.)

To obtain curvature constraints using this method, we perform a Markov Chain Monte Carlo (MCMC) analysis to simultaneously fit $\omegak$, all of the $\{w_i\}$, and several other cosmological parameters to simulated HIRAX and cosmic variance-limited data. The errors on these data were obtained using a Fisher matrix-based likelihood that will be described in Sect.~\ref{section:forecasts}. We also include the Planck constraints on the distance to last scattering, $D_A(z_*)$, to provide a high-redshift anchor point. We use the {\tt emcee} affine-invariant ensemble sampler implemented by \cite{2013PASP..125..306F} to run the MCMC, and then marginalise over all $w_i$ values (and other cosmological parameters) to obtain the marginal distribution for $\omegak$.

\begin{figure}
\centering
\includegraphics[width=\columnwidth]{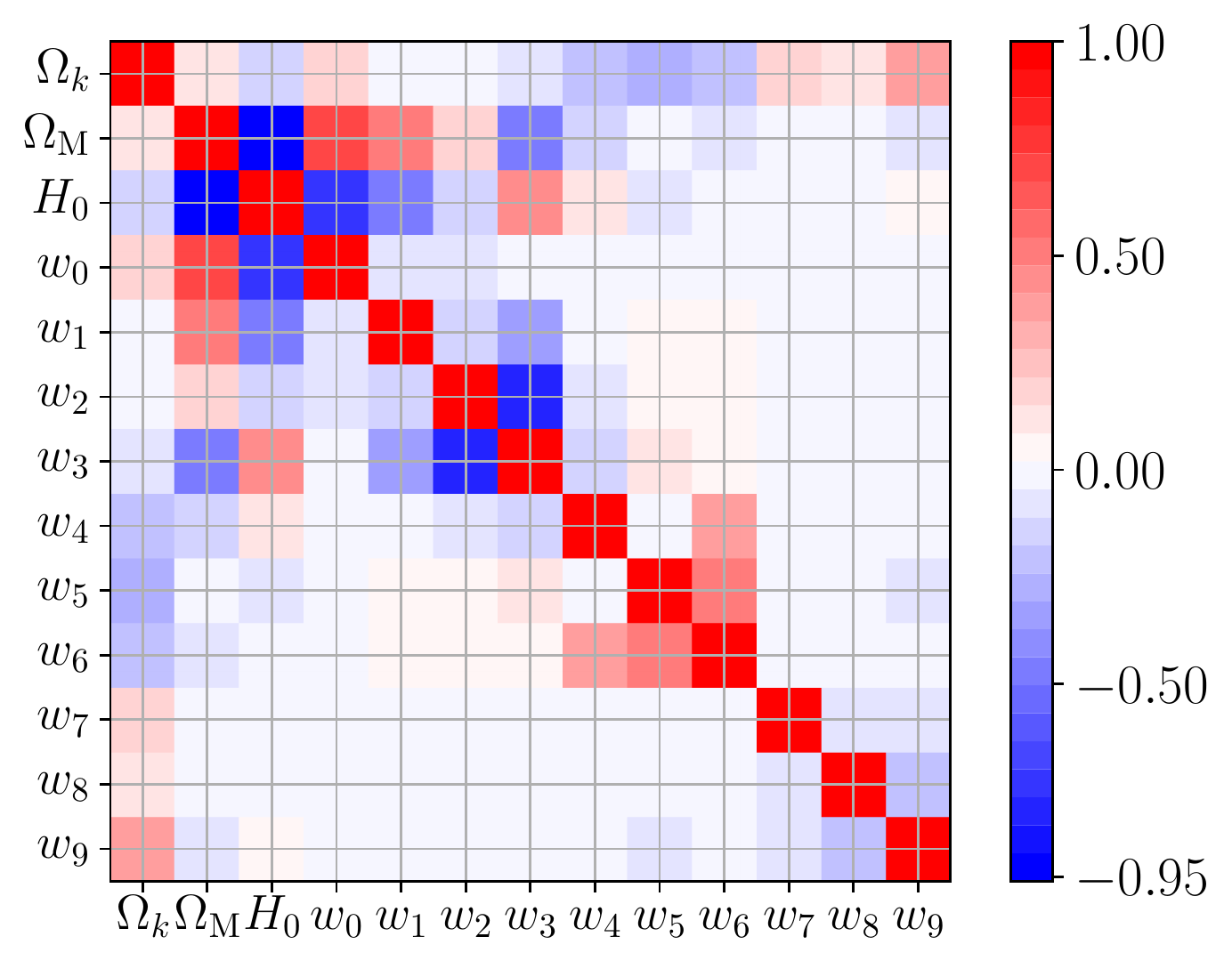}
\vspace{-1em}
\caption{Parameter correlation matrix for HIRAX, for a piecewise dark energy EOS with $10$ bins. The correlation matrix was estimated from the MCMC posteriors.}
\label{fig:correlations}
\end{figure}

Fig.~\ref{fig:correlations} shows the correlations between $\omegak$ and the marginalised parameters for an example MCMC run with 10 EOS bins. For a sufficiently large number of bins, the correlation between $\omegak$ and any individual $w_i$ is relatively mild, but remains non-negligible. The $w_i$ values themselves can be very strongly correlated with one another, however.
%Fig.~\ref{fig:PWCw} shows a set of $w_i$ samples for HIRAX with $\Nde = 10$, coloured by the posterior density $p(w)$ in each bin, where $p(w)$ is normalised to integrate to unity on the flat prior range.
%The resulting correlation matrix for HIRAX is shown in Fig.~\ref{fig:correlations}.

%%%%%%%%%%%%%%%%%%%%%%%%%%%%%%%%%%%%%%%%%%%%%%%%%%
\subsection{Series expansion of dark energy}
\label{section:series}
In order to better compare our results to the literature, we also derive curvature constraints using the common EOS parametrization
%\begin{equation}
$w(z) = w_0 + w_a\frac{z}{(1+z)}$,
%\end{equation}
which gives a dark energy density of
\begin{equation}
\Omega_\mathrm{DE}(z) = \Omega_{\mathrm{DE},0}\, \exp \bigl( -3w_a z / (1+z)\bigr) (1+z)^{3(1+w_0+w_a)}.
\end{equation}
This parameterisation is quite restrictive in that it cannot reproduce the curvature degeneracy in expansion rate data, and so constraints derived using this model will be model dependent.
As in the previous section, we perform an MCMC analysis to find the posterior for $\omegak$, marginalising over all other parameters including $w_0$ and $w_a$.

%%%%%%%%%%%%%%%%%%%%%%%%%%%%%%%%%%%%%%%%%%%%%%%%%%

%%%%%%%%%%%%%%%%%%%%%%%%%%%%%%%%%%%%%%%%%%%%%%%%%%

\section{Forecasts for IM experiments}
\label{section:forecasts}

We now present forecasts for the precision of the model-independent curvature tests that could be performed with forthcoming IM experiments, using HIRAX as an example. To establish the maximum precision of this technique, we also consider two `futuristic' cosmic variance-limited (i.e. thermal noise-free) surveys, with the same array configuration as HIRAX, but redshift ranges of $z = 0.3-3$ (CV1) and $2-5$ (CV2). Such a survey could in principle be accomplished by the future SKA phase 2 (see e.g. \citealt{2015aska.confE..19S}). Note that we do not extend CV1 all the way to $z=0$ as it becomes sensitive only to non-linear scales at low redshifts. For HIRAX we assume the following parameters: frequency resolution, $\delta \nu=0.4$\,MHz, total bandwidth $\Delta\nu = 400$\,MHz, with $\nu_\text{min}=400$\,MHz (giving a redshift range of $0.8$ to $2.5$), system temperature $T_{\rm sys}=50$\,K, the total integration time is assumed to be $t_\mathrm{int}=1$\,yr ($2$\,yr), sky fraction $f_{\rm sky} = 0.25$ ($0.5$), dish diameter $D_\mathrm{dish}=6$\,m and the number of dishes $N_\mathrm{dish}=1024$. Unless stated otherwise we always refer to HIRAX with $t_\mathrm{int}=1$\,yr and $f_{\rm sky} = 0.25$. The dishes are assumed closely packed, arranged in a square grid with $1$\,m of space in between. The CV-limited surveys share the same baseline distribution, but cover the redshift ranges from above and have $f_{\rm sky} = 1$ with no thermal noise.

We begin by assuming that the example surveys will perform measurements of the full anisotropic power spectrum, decomposing it in the radial and transverse directions to obtain constraints on $H(z)$ and $D_A(z)$ respectively. Using the 21cm IM Fisher forecasting code from \cite{2015ApJ...803...21B} and the specifications of the respective surveys,
we obtain covariance matrices for $\{ H(z_j), D_A(z_j) ;\mathrm{~} j = 1 ... N\}$ in a series of $N$ redshift bins $\{z_j\}$ set by the experiment frequency resolution. We impose a non-linear cutoff scale at $z=0$ of $k_\mathrm{NL,0} = 0.2$ Mpc${}^{-1}$, which evolves with redshift according to the results from \cite{Smith:2002dz}, $k_\mathrm{MAX} = k_\mathrm{NL,0}(1+z)^{2/(2+n_s)}$, where $n_s$ is the spectral index of primordial scalar perturbations. In Section \ref{section:results} we examine the dependence of our results on $k_\mathrm{NL,0}$ and compare to the more conservative choice of $k_\mathrm{NL,0} = 0.14$ Mpc${}^{-1}$.

\begin{figure}
\vspace{-1em}
\includegraphics[width=1.\columnwidth]{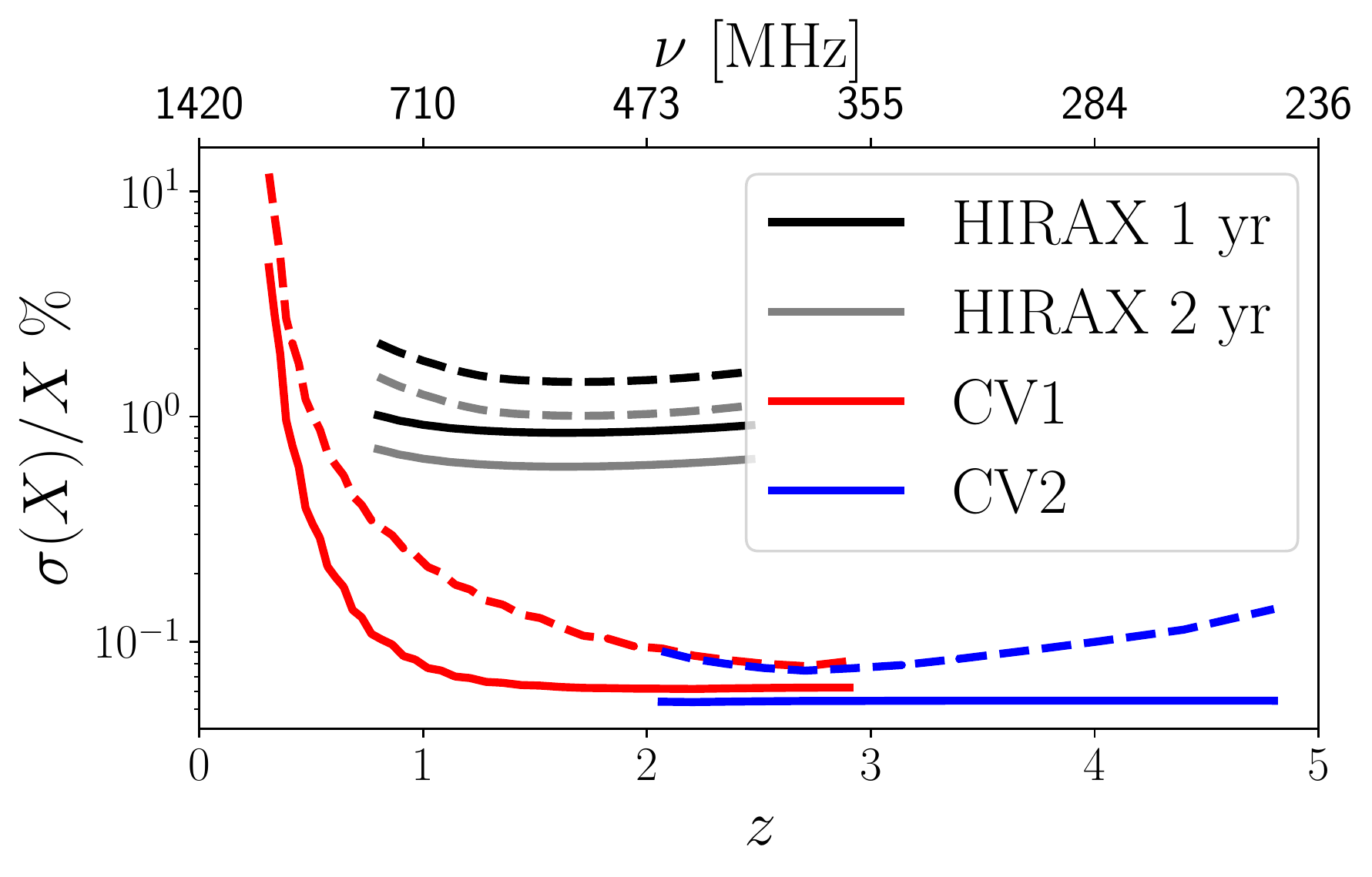}
\vspace{-2em}
\caption{Forecast fractional errors on $H$ (solid lines) and $D_A$ (dashed lines) for HIRAX and the two CV-limited surveys. We assumed a constant frequency binning with $20$ MHz width. The $D_A$ constraints are slightly worse due to (angular) beam smoothing.}
\label{fig:DA_H}
\end{figure}

The form of each covariance matrix is
\begin{equation}
\mathrm{C}(z_j) =
\begin{pmatrix}
\sigma_{D_\mathrm A}^2(z_j)& \sigma_\mathrm{D_A,H}(z_j) \\
\sigma_\mathrm{D_A,H}(z_j) & \sigma_\mathrm {H}^2(z_j)
\end{pmatrix}.
\end{equation}
In obtaining these, we have marginalized over redshift-space distortions (i.e. the growth rate and bias in each bin), as well as the non-linear scale $\sigma_\mathrm{NL}$. The measurements are assumed to be uncorrelated between bins and, optimistically, we have neglected foreground contamination. Instrumental noise, including a realistic baseline distribution, is included in the Fisher matrix calculation however. The errors on $D_A$ and $H$ are shown in Figure \ref{fig:DA_H}.

These covariance matrices, plus the fiducial cosmology, provide the input data of the next step, which is to perform an MCMC analysis to extract the curvature, using each of the model-independent methods described above. The cosmological parameters that we sample are $\Theta = \{\omegak, w_i, \Omega_\mathrm{M}, H_0 \}$, where $i = 1...\Nde$. For simplicity we set the mock data equal to the fiducial functions:
%\begin{equation}
%\label{eq:nosampling}
$D_j \equiv D_A(z_j,\Theta_\mathrm{fid})$ {~and~} $H_j\equiv H(z_j,\Theta_\mathrm{fid}),$
%\end{equation}
where $\Theta_\mathrm{fid}$ is the fiducial set of cosmological parameters, i.e. we do not add noise to the fiducial data vector.

%\begin{figure}
%\includegraphics[width=0.9\columnwidth]{wPWC-eps-converted-to.pdf}
%\caption{Marginal posterior distributions for the piecewise $w(z)$, coloured by probability density $p(w)$, where in each bin we normalised $\int_{-3}^2 p(w)dw = 1$. Except for four bins around $z\sim 1$, the $w_i$ are unconstrained.
%}
%\label{fig:PWCw}
%\end{figure}

Using the definitions $\mu_j=(D_j,H_j)$ and $\xi_j = (D_\mathrm A(z_j, \Theta) , H(z_j, \Theta))$ and omitting additive constants, we can write the log-likelihood function for the MCMC analysis as
$\log\mathcal{L} = -\frac{1}{2}\sum_{j=1}^{N} (\xi_j - \mu_j)^\mathrm T \mathrm{C}^{-1}(z_j) (\xi_j - \mu_j).$

We use only flat priors, with ranges: $w_i\in\interval{-3}{2}$, $\omegak + \Omega_\mathrm M \in\interval{0}{1}$, $\Omega_\mathrm M \in\interval{0}{1}$,
$\omegak\in\interval{-0.5}{0.5}$,
$h\in\interval{0.1}{1}$. We also include CMB angular diameter distance data from Planck as additional data points in the MCMC analysis. The standard deviation of the sound horizon at recombination $r_*$, of the acoustic peak scale $\Theta_*$, and of the mean redshift of the last scattering surface $z_*$ are taken from \cite{2016A&A...594A..13P}. With $\Theta_* = (r_*/D_{A,*})/(1+z_*)$, standard error propagation gives $\sigma(D_{A,*}) = 0.044$ Mpc at $z_* = 1090.09$. The contribution of neutrinos and radiation to the mean energy density is assumed to be fixed at $\Omega_\mathrm{rad} = 9.13 \times 10 ^{-5}$, also from the Planck results.

%%%%%%%%%%%%%%%%%%%%%%%%%%%%%%%%%%%%%%%%%%%%%%%%%%

\subsection{Convergence of the curvature constraints}

\begin{figure}
%\vspace{-1em}
\includegraphics[width=0.9\columnwidth]{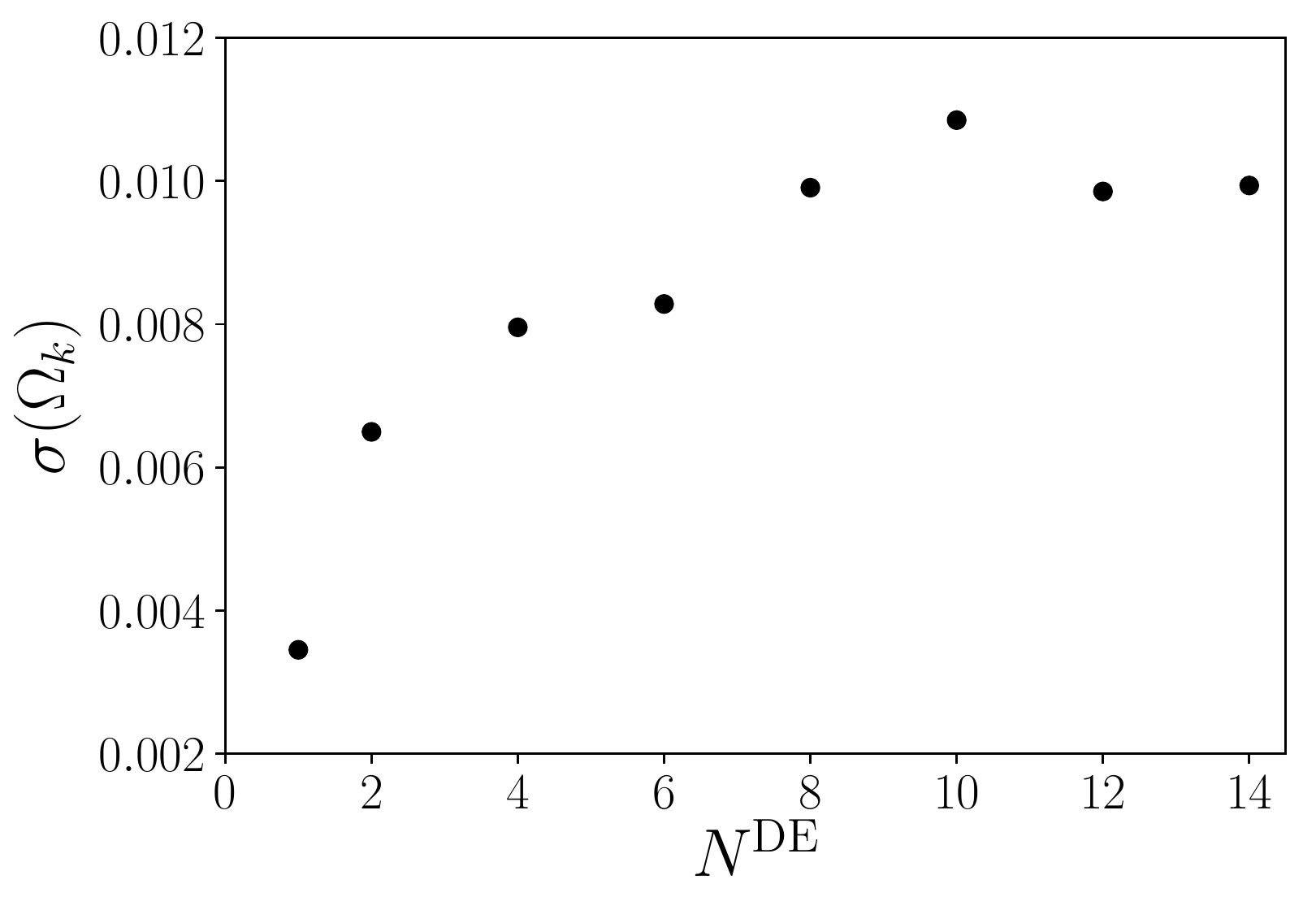}
\caption{Forecast $1\sigma$ constraints on curvature for HIRAX as a function of the number of DE equation of state bins that are marginalised over, $\Nde$. The constraints converge for $\Nde \gtrsim 8$.}
\label{fig:convergence}
\end{figure}

The constraints on the cosmological parameters yielded by a given experiment generally depend on the choice of dark energy model. In our case, the constraints depend on the choice of the binning of $w(z)$, and on the number of bins $\Nde$. In Figure~\ref{fig:convergence}, we show the behavior of $\sigma(\omegak)$ as a function of $\Nde$ for a binning that is equally-spaced in scale factor, $a$. For a given experiment the curvature constraints converge once the number of bins is large enough, suggesting that the dependence on the particular form of the DE equation of state model has been removed after this point. We find convergence for $\Nde \gtrsim 8$ for SDSS, $10$ for HIRAX, and $16$ for CV1 and CV2.

%%%%%%%%%%%%%%%%%%%%%%%%%%%%%%%%%%%%%%%%%%%%%%%%%%

\subsection{Results}
\label{section:results}

\begin{figure}
\centering
\vspace{-1em}
\includegraphics[width=1.05\columnwidth]{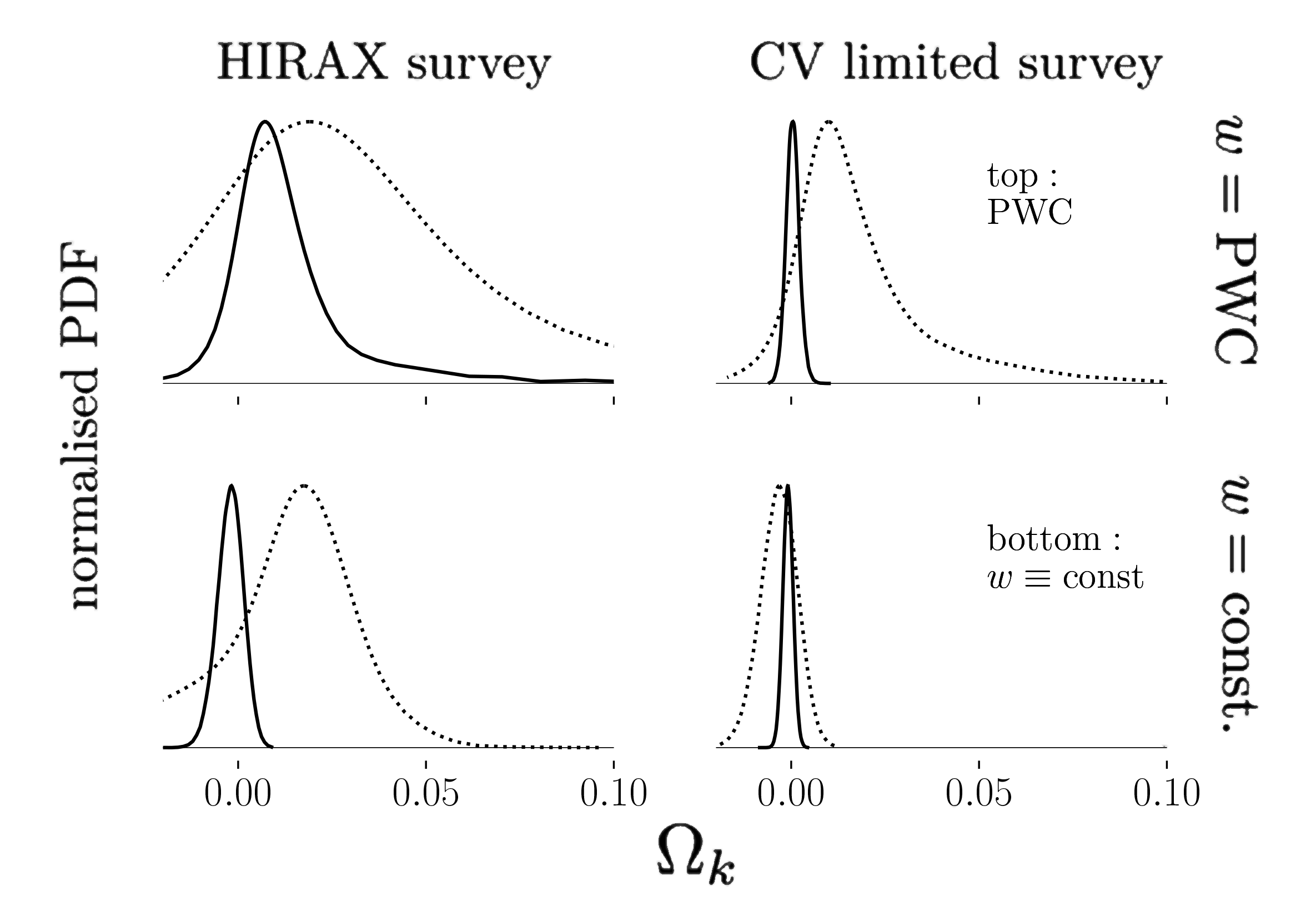}
\vspace{-2em}
% \vspace{1em}
\caption{The (rescaled) posterior distributions of $\omegak$ for HIRAX (left) and CV1 (right). The bottom row is for a dark energy model with constant $w$, and the top row is for a model with 10 bins (HIRAX) and 16 bins (CV1) in the piecewise constant parametrisation of the EOS. Solid lines correspond to the full analysis, while the dotted lines use $D_A$ measurements only.}
\label{fig:posteriors}
\end{figure}

Table \ref{tab:res} lists the 68\% errors on $\omegak$ for different methods of marginalising out dark energy. While the entries for Planck and SDSS use actual data, we used Fisher forecasts for the planned/hypothetical surveys HIRAX, CV1, and CV2, as described above. The posteriors for HIRAX and CV1 for the $w={\rm const.}$ and piecewise constant models are shown in Figure~\ref{fig:posteriors}, where we also compare the full $H$ and $D_A$ constraints to what would be measured using $D_A$ data only.

One can clearly see how much the degeneracy depends on the choice of the dark energy model. While measurements of $D_A$ alone are enough to constrain $\omegak$ reasonably well for constant $w(z) \equiv w$, this is not the case for the much more general piecewise constant model, which requires $H(z)$ measurements as well to reach an appreciable level of precision. Even for a model with an arbitrarily large number of EOS bins, though, combining $H(z)$ and $D_A(z)$ breaks the curvature-dark energy degeneracy and allows for constraints as good as $\sigma(\omegak) \approx 2\times 10^{-3}$ for the cosmic variance-limited surveys. The $2$\, yr survey for HIRAX is a factor of $\sqrt{2}$ better than $1$\, yr in the avoidance analysis. When the other methods are used the improvement is somewhat weaker, tightening the constraints by about 30\% in the piecewise DE model. CV1 slightly outperforms the higher redshift CV2, even though it covers higher redshifts that are less sensitive to dark energy. This is due to the array setup we assumed being designed for BAO detection in the lower redshift range, so its resolution is worse at higher redshift.

\begin{figure}
%\centering
\hspace{-1.1em}
\includegraphics[width=1.05\columnwidth]{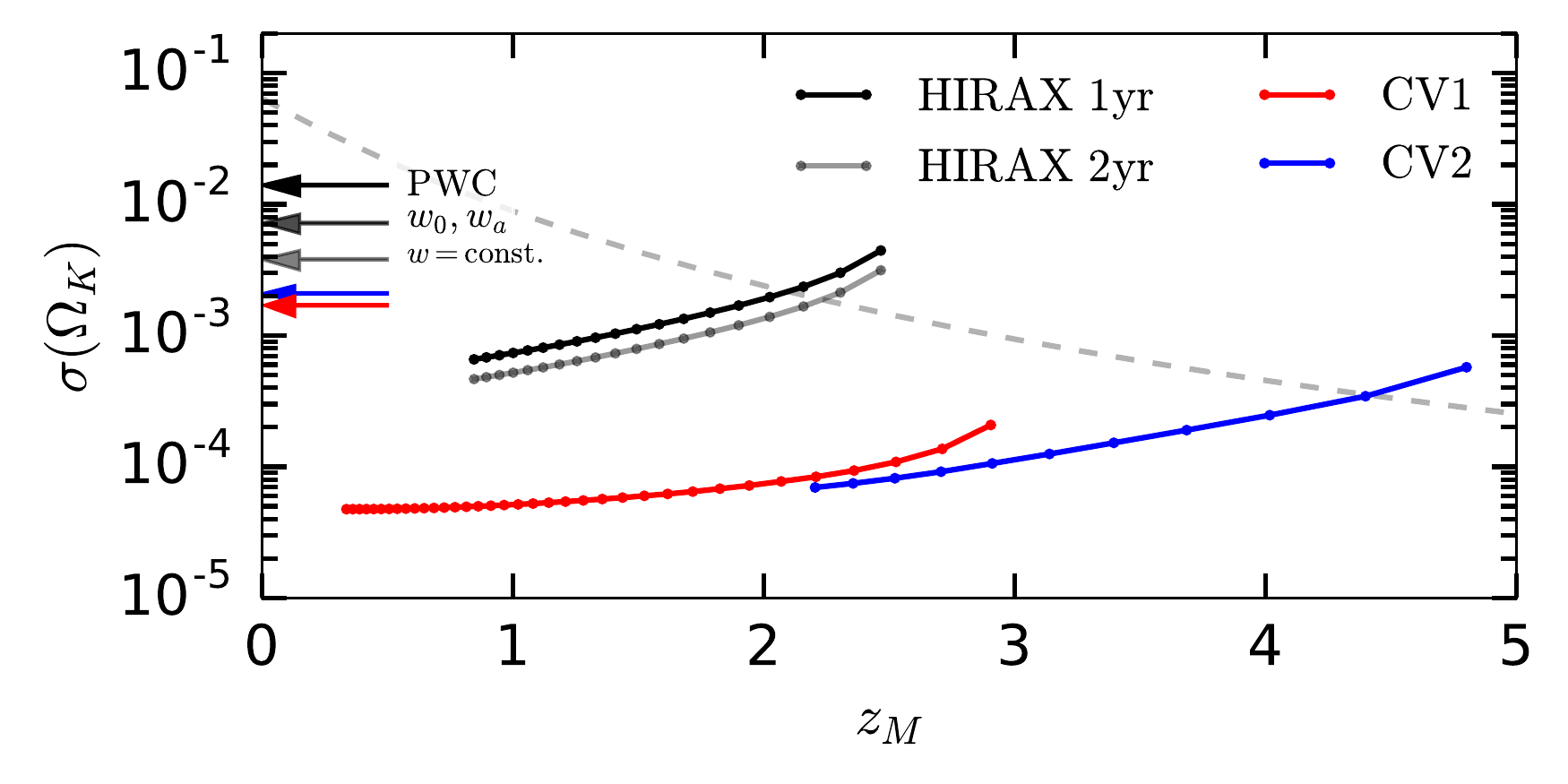}
\vspace{-1.5em}
\caption{Constraints on $\omegak$ as a function of cutoff redshift, $z_M$, for the avoidance method. The dashed grey line shows $\Delta \omegak$ in a $w=-1$ model (c.f. Fig.~\ref{fig:kbias}). The horizontal bars mark the 68\% CL limits for HIRAX for all of the other methods (black/grey), and for just the PWC method for CV1 and CV2 (red/blue).}
\label{fig:avoidance}
\end{figure}

The avoidance method, on the other hand, can yield constraints at the $\sim 10^{-4}$ level in the cosmic variance-limited case, with errorbars a factor of a few smaller than the piecewise constant model for HIRAX. This improvement in precision must be balanced against the potential bias that is introduced by simply ignoring dark energy, as illustrated in Fig.~\ref{fig:kbias}. Making the expected bias smaller than the errorbars will require either the reintroduction of a (possibly much simpler) dark energy model, or a higher redshift cutoff $z_M$. The effect of changing the cutoff is shown in Fig.~\ref{fig:avoidance}; we see that $z_M \gtrsim 2$ is sufficient for the HIRAX measurement to not be dominated by the bias for $w=-1$, while $z_M \gtrsim 4.5$ is needed for CV2.

Note that all of these results can depend on the choice of nonlinear cutoff scale, $k_\mathrm{MAX}$. If a more conservative value of $k_\mathrm{NL,0} = 0.14$ Mpc${}^{-1}$ is chosen, we find $\omegak = (9.6^{\plus 14.1}_{\minus 7.5}) \times 10^{-3}$ (68\% CL) for HIRAX with the piecewise DE model. This is consistent with the results for the more optimistic choice of $k_\mathrm{NL,0} = 0.2$ Mpc${}^{-1}$ that was used throughout this work.

\begin{table}
%\scriptsize
\begin{center}
\begin{tabular}{| l | l | l | l | l | }
%\hline
 & Avoidance &  $w\equiv \mathrm{const}$ & $w_0w_a{}$ & Piecewise \\ \hline \tvspace
	Planck  & --- & $-52^{\plus49}_{\minus 55}$ & --- &  ---  \\ \tvspace
   SDSS  & --- & $+39^{\plus 29}_{\minus 70}$ & --- & $+76^{\plus 65}_{\minus 50} $  \\ \tvspace
   HIRAX $1$ yr & $0.0^{\plus 2.0}_{\minus 2.0}$& $-2.0^{\plus 3.3}_{\minus 3.6}$ & $-1.3^{\plus 6.2}_{\minus 7.0}$ & $+9.35^{\plus 13.9}_{\minus 7.76} $ \\ \tvspace
   HIRAX $2$ yr & $0.0^{\plus 1.4}_{\minus 1.4}$& $-2.0^{\plus 2.8}_{\minus 2.9}$ & $-2.0^{\plus 5.3}_{\minus 6.0}$ & $7.6^{\plus 10.3}_{\minus 6.6} $ \\ \tvspace
   CV1 & $0.0^{\plus 0.07}_{\minus 0.07} $ & $-0.9^{\plus1.4}_{\minus 1.4}$ &$-0.9^{\plus1.4}_{\minus 1.4}$ & $+0.4^{\plus 1.7}_{\minus 1.7}$ \\ \tvspace
   CV2 & $0.0^{\plus 0.07}_{\minus 0.07}$ & $-1.1^{\plus 1.6}_{\minus 1.6}$ &$-1.1^{\plus1.6}_{\minus 1.6}$ &  $-0.1^{\plus 2.1}_{\minus 1.9} $ %\\%\hline
\normalsize
\end{tabular}%}
\vspace{0.5em}
\caption{Measured (Planck/SDSS) and forecast (HIRAX/CV) constraints on $\omegak$ at $68\%$ confidence (in units of $10^{-3}$), for different dark energy models and analysis methods. For the piecewise constant model, $10$ bins were assumed for HIRAX with both $1$ and $2$\,yr integration, $8$ for SDSS, and $16$ for CV1 and CV2. Planck CMB distance constraints are included in all of these results. For the avoidance approach, a cutoff of $z_M = 2$ was imposed; the expected bias $\Delta \omegak$ was not included in the errors, but is large (see Fig.~\ref{fig:avoidance}).}
\label{tab:res}
\end{center}
\end{table}

%%%%%%%%%%%%%%%%%%%%%%%%%%%%%%%%%%%%%%%%%%%%%%%%%%
\vspace{-1em}
\section{Conclusions}

In this work, we showed what constraints can be achieved on the curvature when assuming a completely generic dark energy model. We started by Fisher forecasting errors on $D_A$ and $H$ for HIRAX and two cosmic variance-limited surveys, then derived curvature constraints in two different ways that were designed to largely avoid any dependence on the chosen dark energy model.

In the first approach, an extension of the one presented in \cite{2006PhRvD..73b3503K}, we made the assumption that dark energy could be neglected above some $z > z_M$, far into the matter-dominated regime. This largely removes the dependence of the curvature constraints on the DE EOS, at the cost of throwing away information at lower redshift. It is able to produce quite tight constraints on $\omegak$ however, and while the method is biased, this can be reduced with an appropriate choice of $z_M$. For a target precision of $\sigma(\omegak) \sim 10^{-4}$, a redshift cut of $z_M \gtrsim 4$ or more would be necessary.

In the second approach, we adopted a piecewise constant parametrization of the dark energy EOS $w(z)$ and used an MCMC method to sample from the posterior of $\omegak$, marginalised over the values of $w(z)$ in each bin. This does not require any low-$z$ information to be discarded, is unbiased in principle, and requires fewer assumptions than the first method. The constraints obtained depend on the choice of binning, especially on the number of bins $\Nde$, but converge once $\Nde$ is high enough ($\sim 10$ for HIRAX).
%Instead of using constraints on cosmological parameters from the CMB, we used the measurement of the CMB distance from Planck as an additional data point in the likelihood calculation. This procedure only allows for meaningful constraints on $\omegak$ if $H(z)$ and $D_A(z)$ measurements are combined, as shown in Figure~\ref{fig:posteriors} (this is not necesary in simpler models, e.g. with $w(z) \equiv \mathrm{const.}$).
Generally, this method produces constraints that are an order of magnitude weaker than the avoidance method, but this is reduced to only a factor of $2-3$ when the bias of the latter is factored in.

In conclusion, neither the avoidance nor non-parametric marginalisation method is able to reach the target precision of $\sigma(\omegak) \sim 10^{-4}$ set by eternal inflation models, at least with the instrumental setup we assumed. A possible exception is if the avoidance method is used at very high redshift, $z_M \gtrsim 5$, where the bias should be significantly reduced.

% Balance columns on last page
\balance

\section*{Acknowledgements}
We are grateful to Tony Walters and Frederik Beaujean for useful discussions. A. Witzemann and MGS acknowledge support from the South African Square Kilometre Array Project and National Research Foundation of South Africa. PB acknowledges support from an appointment to the NASA Postdoctoral Program at the Jet Propulsion Laboratory, California Institute of Technology, administered by Universities Space Research Association under contract with NASA. CC was supported by STFC Consolidated Grant ST/P000592/1. A. Weltman is supported by the South African Research Chairs Initiative of the Department of Science and Technology and the National Research Foundation of South Africa. She is also grateful for support from the Institute for Advanced study, the Flatiron Institute and the Princeton University Astrophysics department. Any opinion, finding and conclusion or recommendation expressed in this material is that of the authors and the NRF does not accept any liability in this regard.

%%%%%%%%%%%%%%%%%%%% REFERENCES %%%%%%%%%%%%%%%%%%

\bibliographystyle{mnras}
\bibliography{HIRAX_DE}

\begin{thebibliography}{}
\makeatletter
\relax
\def\mn@urlcharsother{\let\do\@makeother \do\$\do\&\do\#\do\^\do\_\do\%\do\~}
\def\mn@doi{\begingroup\mn@urlcharsother \@ifnextchar [ {\mn@doi@}
  {\mn@doi@[]}}
\def\mn@doi@[#1]#2{\def\@tempa{#1}\ifx\@tempa\@empty \href
  {http://dx.doi.org/#2} {doi:#2}\else \href {http://dx.doi.org/#2} {#1}\fi
  \endgroup}
\def\mn@eprint#1#2{\mn@eprint@#1:#2::\@nil}
\def\mn@eprint@arXiv#1{\href {http://arxiv.org/abs/#1} {{\tt arXiv:#1}}}
\def\mn@eprint@dblp#1{\href {http://dblp.uni-trier.de/rec/bibtex/#1.xml}
  {dblp:#1}}
\def\mn@eprint@#1:#2:#3:#4\@nil{\def\@tempa {#1}\def\@tempb {#2}\def\@tempc
  {#3}\ifx \@tempc \@empty \let \@tempc \@tempb \let \@tempb \@tempa \fi \ifx
  \@tempb \@empty \def\@tempb {arXiv}\fi \@ifundefined
  {mn@eprint@\@tempb}{\@tempb:\@tempc}{\expandafter \expandafter \csname
  mn@eprint@\@tempb\endcsname \expandafter{\@tempc}}}

\bibitem[\protect\citeauthoryear{{Aslanyan} \& {Easther}}{{Aslanyan} \&
  {Easther}}{2015}]{2015PhRvD..91l3523A}
{Aslanyan} G.,  {Easther} R.,  2015, \mn@doi [\prd]
  {10.1103/PhysRevD.91.123523}, \href
  {http://adsabs.harvard.edu/abs/2015PhRvD..91l3523A} {91, 123523}

\bibitem[\protect\citeauthoryear{{Bucher}, {Goldhaber}  \& {Turok}}{{Bucher}
  et~al.}{1995}]{1995PhRvD..52.3314B}
{Bucher} M.,  {Goldhaber} A.~S.,   {Turok} N.,  1995, \mn@doi [\prd]
  {10.1103/PhysRevD.52.3314}, \href
  {http://adsabs.harvard.edu/abs/1995PhRvD..52.3314B} {52, 3314}

\bibitem[\protect\citeauthoryear{Bull, Camera, Raccanelli, Blake, Ferreira,
  Santos  \& Schwarz}{Bull et~al.}{2015a}]{Bull:2015nra}
Bull P.,  Camera S.,  Raccanelli A.,  Blake C.,  Ferreira P.~G.,  Santos M.~G.,
    Schwarz D.~J.,  2015a, in {PoS AASKA14 (2015) 024}.  (\mn@eprint {arXiv}
  {1501.04088})

\bibitem[\protect\citeauthoryear{{Bull}, {Ferreira}, {Patel}  \&
  {Santos}}{{Bull} et~al.}{2015b}]{2015ApJ...803...21B}
{Bull} P.,  {Ferreira} P.~G.,  {Patel} P.,   {Santos} M.~G.,  2015b, \mn@doi
  [\apj] {10.1088/0004-637X/803/1/21}, \href
  {http://adsabs.harvard.edu/abs/2015ApJ...803...21B} {803, 21}

\bibitem[\protect\citeauthoryear{Chang, Pen, Peterson  \& McDonald}{Chang
  et~al.}{2008}]{Chang:2007xk}
Chang T.-C.,  Pen U.-L.,  Peterson J.~B.,   McDonald P.,  2008, \mn@doi [Phys.
  Rev. Lett.] {10.1103/PhysRevLett.100.091303}, 100, 091303

\bibitem[\protect\citeauthoryear{{Clarkson}, {Cort{\^e}s}  \&
  {Bassett}}{{Clarkson} et~al.}{2007}]{2007JCAP...08..011C}
{Clarkson} C.,  {Cort{\^e}s} M.,   {Bassett} B.,  2007, \mn@doi [\jcap]
  {10.1088/1475-7516/2007/08/011}, \href
  {http://adsabs.harvard.edu/abs/2007JCAP...08..011C} {8, 011}

\bibitem[\protect\citeauthoryear{{Cornish}, {Spergel}  \& {Starkman}}{{Cornish}
  et~al.}{1996}]{1996PhRvL..77..215C}
{Cornish} N.~J.,  {Spergel} D.~N.,   {Starkman} G.~D.,  1996, \mn@doi [Physical
  Review Letters] {10.1103/PhysRevLett.77.215}, \href
  {http://adsabs.harvard.edu/abs/1996PhRvL..77..215C} {77, 215}

\bibitem[\protect\citeauthoryear{{Foreman-Mackey}, {Hogg}, {Lang}  \&
  {Goodman}}{{Foreman-Mackey} et~al.}{2013}]{2013PASP..125..306F}
{Foreman-Mackey} D.,  {Hogg} D.~W.,  {Lang} D.,   {Goodman} J.,  2013, \mn@doi
  [\pasp] {10.1086/670067}, \href
  {http://adsabs.harvard.edu/abs/2013PASP..125..306F} {125, 306}

\bibitem[\protect\citeauthoryear{{Guth} \& {Nomura}}{{Guth} \&
  {Nomura}}{2012}]{2012PhRvD..86b3534G}
{Guth} A.~H.,  {Nomura} Y.,  2012, \mn@doi [\prd] {10.1103/PhysRevD.86.023534},
  \href {http://adsabs.harvard.edu/abs/2012PhRvD..86b3534G} {86, 023534}

\bibitem[\protect\citeauthoryear{{Hlozek}, {Cort{\^e}s}, {Clarkson}  \&
  {Bassett}}{{Hlozek} et~al.}{2008}]{2008GReGr..40..285H}
{Hlozek} R.,  {Cort{\^e}s} M.,  {Clarkson} C.,   {Bassett} B.,  2008, \mn@doi
  [General Relativity and Gravitation] {10.1007/s10714-007-0548-6}, \href
  {http://adsabs.harvard.edu/abs/2008GReGr..40..285H} {40, 285}

\bibitem[\protect\citeauthoryear{Huterer \& Peiris}{Huterer \&
  Peiris}{2007}]{Huterer:2006mv}
Huterer D.,  Peiris H.~V.,  2007, \mn@doi [Phys. Rev.]
  {10.1103/PhysRevD.75.083503}, D75, 083503

\bibitem[\protect\citeauthoryear{Kleban \& Schillo}{Kleban \&
  Schillo}{2012}]{Kleban2012}
Kleban M.,  Schillo M.,  2012, \jcap, 2012, 029

\bibitem[\protect\citeauthoryear{{Knox}}{{Knox}}{2006}]{2006PhRvD..73b3503K}
{Knox} L.,  2006, \mn@doi [\prd] {10.1103/PhysRevD.73.023503}, \href
  {http://adsabs.harvard.edu/abs/2006PhRvD..73b3503K} {73, 023503}

\bibitem[\protect\citeauthoryear{{Kovetz} et~al.,}{{Kovetz}
  et~al.}{2017}]{2017arXiv170909066K}
{Kovetz} E.~D.,  et~al., 2017, preprint, \href
  {http://adsabs.harvard.edu/abs/2017arXiv170909066K} {} (\mn@eprint {arXiv}
  {1709.09066})

\bibitem[\protect\citeauthoryear{Leonard, Bull  \& Allison}{Leonard
  et~al.}{2016}]{Leonard:2016evk}
Leonard C.~D.,  Bull P.,   Allison R.,  2016, \mn@doi [Phys. Rev.]
  {10.1103/PhysRevD.94.023502}, D94, 023502

\bibitem[\protect\citeauthoryear{Marsh, Bull, Ferreira  \& Pontzen}{Marsh
  et~al.}{2014}]{Marsh:2014xoa}
Marsh D. J.~E.,  Bull P.,  Ferreira P.~G.,   Pontzen A.,  2014, \mn@doi [Phys.
  Rev.] {10.1103/PhysRevD.90.105023}, D90, 105023

\bibitem[\protect\citeauthoryear{{Nesseris} \& {Sapone}}{{Nesseris} \&
  {Sapone}}{2014}]{2014PhRvD..90f3006N}
{Nesseris} S.,  {Sapone} D.,  2014, \mn@doi [\prd]
  {10.1103/PhysRevD.90.063006}, \href
  {http://adsabs.harvard.edu/abs/2014PhRvD..90f3006N} {90, 063006}

\bibitem[\protect\citeauthoryear{{Newburgh} et~al.,}{{Newburgh}
  et~al.}{2016}]{2016SPIE.9906E..5XN}
{Newburgh} L.~B.,  et~al., 2016, in Ground-based and Airborne Telescopes VI. p.
  99065X (\mn@eprint {arXiv} {1607.02059}), \mn@doi{10.1117/12.2234286}

\bibitem[\protect\citeauthoryear{{Planck Collaboration XIII}}{{Planck
  Collaboration XIII}}{2016}]{2016A&A...594A..13P}
{Planck Collaboration XIII} 2016, \mn@doi [\aap] {10.1051/0004-6361/201525830},
  \href {http://adsabs.harvard.edu/abs/2016A%26A...594A..13P} {594, A13}

\bibitem[\protect\citeauthoryear{Raveri, Bull, Silvestri  \& Pogosian}{Raveri
  et~al.}{2017}]{Raveri:2017qvt}
Raveri M.,  Bull P.,  Silvestri A.,   Pogosian L.,  2017, arXiv:1703.05297

\bibitem[\protect\citeauthoryear{{S{\'a}nchez} et~al.,}{{S{\'a}nchez}
  et~al.}{2017}]{2017MNRAS.464.1493S}
{S{\'a}nchez} A.~G.,  et~al., 2017, \mn@doi [\mnras] {10.1093/mnras/stw2495},
  \href {http://adsabs.harvard.edu/abs/2017MNRAS.464.1493S} {464, 1493}

\bibitem[\protect\citeauthoryear{{Santos} et~al.,}{{Santos}
  et~al.}{2015}]{2015aska.confE..19S}
{Santos} M.,  et~al., 2015, Advancing Astrophysics with the Square Kilometre
  Array (AASKA14), \href {http://adsabs.harvard.edu/abs/2015aska.confE..19S}
  {p.~19}

\bibitem[\protect\citeauthoryear{Smith et~al.,}{Smith
  et~al.}{2003}]{Smith:2002dz}
Smith R.~E.,  et~al., 2003, \mn@doi [Mon. Not. Roy. Astron. Soc.]
  {10.1046/j.1365-8711.2003.06503.x}, 341, 1311

\bibitem[\protect\citeauthoryear{Takada \& Dore}{Takada \&
  Dore}{2015}]{Takada:2015mma}
Takada M.,  Dore O.,  2015, \mn@doi [Phys. Rev.] {10.1103/PhysRevD.92.123518},
  D92, 123518

\bibitem[\protect\citeauthoryear{Vardanyan, Trotta  \& Silk}{Vardanyan
  et~al.}{2009}]{Vardanyan:2009ft}
Vardanyan M.,  Trotta R.,   Silk J.,  2009, \mn@doi [Mon. Not. Roy. Astron.
  Soc.] {10.1111/j.1365-2966.2009.14938.x}, 397, 431

\bibitem[\protect\citeauthoryear{Villaescusa-Navarro, Alonso  \&
  Viel}{Villaescusa-Navarro et~al.}{2017}]{Villaescusa-Navarro:2016kbz}
Villaescusa-Navarro F.,  Alonso D.,   Viel M.,  2017, \mn@doi [Mon. Not. Roy.
  Astron. Soc.] {10.1093/mnras/stw3224}, 466, 2736

\makeatother
\end{thebibliography}

%%%%%%%%%%%%%%%%%%%%%%%

% Don't change these lines
%\bsp	% typesetting comment
\label{lastpage}
\end{document}